\documentclass[pre,aps,twocolumn,showpacs,floatfix,10pt]{revtex4-1} 
\usepackage{epsfig}
\usepackage{psfrag}
\usepackage{amsmath}
\usepackage{amsfonts}
\usepackage{amssymb}

\usepackage[normalem]{ulem} 

\usepackage{graphicx}

\usepackage{color}

\definecolor{RED}{rgb}{1,0,0}
\definecolor{BLUE}{rgb}{0,0,1}
\definecolor{DARKBLUE}{rgb}{0,0,0.4}
\definecolor{GREEN}{rgb}{0,0.5,0}
\definecolor{DARKGREEN}{rgb}{0,0.3,0}
\definecolor{MAGENTA}{rgb}{1,0,1}
\definecolor{PURPLE}{rgb}{0.5,0,0.5}

\newcommand{\vc}[0]{\boldsymbol}
\newcommand{\mtx}[0]{\mathbf}

\newcommand{\norm}[1]{\left|\left| #1 \right|\right|}
\newcommand{\reals}[0]{\mathbb{R}}

\newif\ifusebibtex
\usebibtexfalse

\newif\ifarxiv 
\arxivtrue

\begin{document}
\title{Critical slowing down and hyperuniformity on approach to jamming}
\author{Steven Atkinson}
\affiliation{Department of Mechanical and Aerospace Engineering, Princeton University, Princeton, New Jersey 08544, USA}
\author{Ge Zhang}
\affiliation{Department of Chemistry, Princeton University, Princeton, New Jersey 08544, USA}
\author{Adam B. Hopkins}
\affiliation{Uniformity Labs, 1600 Adams Drive, Suite 104, Menlo Park, CA 94025}
\author{Salvatore Torquato}
\email{torquato@princeton.edu}
\affiliation{Department of Chemistry, Department of Physics, Princeton Center for Theoretical Science, Program of Applied and Computational Mathematics, Princeton Institute for the Science and Technology of Materials, Princeton University, Princeton, New Jersey 08544, USA}
\date{\today}


\begin{abstract}
Hyperuniformity characterizes a state of matter that is poised at a critical point at which density or volume-fraction fluctuations are anomalously suppressed at infinite wavelengths.
Recently, much attention has been given to the link between strict jamming (mechanical rigidity) and (effective or exact) hyperuniformity in frictionless hard-particle packings.
However, in doing so, one must necessarily study very large packings in order to access the long-ranged behavior and to ensure that the packings are truly jammed.
We modify the rigorous linear programming method of Donev et al.\ [J.\ Comp.\ Phys.\ {\bf 197}, 139 (2004)] in order to test for jamming in putatively collectively and strictly jammed packings of hard-disks in two dimensions.
We show that this rigorous jamming test is superior to standard ways to ascertain jamming, including the so-called ``pressure-leak'' test. We find that various standard packing protocols struggle to reliably create packings that are jammed for even modest system sizes of $N \approx 10^3$  bidisperse disks in two dimensions; importantly, these packings have a high reduced pressure that persists over extended amounts of time, meaning that they {\it appear} to be jammed by conventional tests, though rigorous jamming tests reveal that they are not.
We present evidence that suggests that deviations from hyperuniformity in putative maximally random jammed (MRJ) packings can in part be explained by a shortcoming of the numerical protocols to generate exactly-jammed configurations as a result of a type of ``critical slowing down'' as the packing's collective rearrangements in configuration space become locally confined by high-dimensional ``bottlenecks'' from which escape is a rare event.
Additionally, various protocols are able to produce packings exhibiting hyperuniformity to different extents, but this is because certain protocols are better able to approach exactly-jammed configurations.
Nonetheless, while one should not generally expect exact hyperuniformity for disordered packings with rattlers, we find that when jamming is ensured, our packings are very nearly hyperuniform, and deviations from hyperuniformity correlate with an inability to ensure jamming, suggesting that strict jamming and hyperuniformity are indeed linked.
This raises the possibility that the ideal MRJ packings have no rattlers.
Our work provides the impetus for the development of packing algorithms that produce large disordered strictly jammed packings that are rattler-free, which is  an outstanding, challenging task.
\end{abstract}
\maketitle



\section{Introduction}

Dense packings of hard (nonoverlapping) spheres in $d$-dimensional Euclidean space $\reals^d$ have been a source of fascination to scientists across the physical and mathematical sciences.  Particle packings have served as simple, yet powerful models for a wide variety of condensed matter systems including liquids, glasses, colloids, particulate composites, granular materials, and biological systems, to name a few \cite{Bernal_1965,Zallen_1983,Cates_1999,Torquato_2001,Torquato_2003_Breakdown,Torquato_2003_Hyperuniform,Liu_2003,Manoharan_2003,Makse_2004,Xu_2005,Zohdi_2006,Majmudar_2007,Gevertz_2008,Laso_2009,Parisi_2010,Schweizer_2010,Liu_2011,Charbonneau_2012,Dagois-Bohy_2012,Gillman_2013,Zohdi_2013,Klatt_2014,Bowles_2015,Chaikin_2015,Schroder-Turk_2015,Chakraborty_2016}.  Of particular interest are mechanically stable or jammed packings \cite{Torquato_2001,Liu_2003,Xu_2005,Parisi_2010,Torquato_2010,Charbonneau_2012,Dagois-Bohy_2012}.  In order to make the notion of jamming rigorous, Torquato et al.\ introduced the following rigorous hierarchical jamming categories for frictionless spheres in $\reals^d$ \cite{Torquato_2001,Torquato_2003_Breakdown}: a {\it locally jammed} packing is one in which no particle may be displaced while all others are fixed in place.  A {\it collectively jammed} packing is one in which no subset of particles may be displaced while fixing the shape of the system boundary.  A {\it strictly jammed} packing is a packing in which no subset of particles may be displaced while allowing volume-preserving deformations of the system boundary \cite{Torquato_2003_Breakdown}.  Thus, all strictly jammed packings are collectively jammed, and all collectively jammed packings are locally jammed (each particle is locally trapped by at least $d+1$ contacting spheres not all in the same hemisphere).  Collectively jammed packings are stable against uniform compression, i.e., their bulk modulus is positive; and strictly jammed packings are additionally stable against shear, implying they also have a positive shear modulus.  In the limit of exact strict jamming, the bulk and shear moduli of hard-particle packings both diverge to infinity \cite{Torquato_2003_Breakdown}.


Torquato and Stillinger have conjectured that any strictly jammed saturated infinite packing of identical spheres is hyperuniform \cite{Torquato_2003_Hyperuniform,foot:infinite}.
A {\it saturated}  packing of hard spheres is one in which there is no space available to add another sphere. Any {\it hyperuniform} point pattern is poised
at a ``critical point'' because it is characterized by an anomalously large suppression of large-scale density fluctuations such that the direct
correlation function is long-ranged \cite{Torquato_2003_Hyperuniform}, which is manifested by a local number variance $\sigma^2(R)$ that grows more slowly
than $R^d$ for a spherical observation window of radius $R$ or, equivalently, a structure factor $S(\vc{k})$ that tends to zero 
as the wavenumber $|\vc{k}|$ tends to zero.  More generally,
for two-phase media, hyperuniformity is manifested by a local volume-fraction variance that decays more rapidly than $R^{-d}$ or, equivalently, by a spectral density ${\tilde \chi}(\vc{k})$ \cite{Torquato_2002,foot:spectral} that tends to zero in the limit $|\vc{k}| \rightarrow 0$ \cite{Zachary_2009}.  To date, there is no known counterexample to this conjecture, notwithstanding a recent study that calls into question the link between jamming and hyperuniformity \cite{Ikeda_2015}.

What is the rationale for such a conjecture? First, we note that a packing can be hyperuniform without being strictly jammed. For example, a honeycomb lattice packing of identical circular disks in two dimensions is only locally jammed but is hyperuniform with positive bulk modulus and zero shear modulus. This example
stresses the importance of the strict jamming constraint.  Indeed, appropriate deformations and compressions of a packing that is only locally or collectively jammed (and hence hypostatic with respect to strict jamming) can lead to a denser strictly jammed packing that is isostatic or hyperstatic \cite{Donev_2004_Jamming}, filling space more uniformly.  We also know that there are infinite periodic packings, such as the triangular lattice of identical circular disks in $\mathbb{R}^2$ and face-centered-cubic lattice in $\mathbb{R}^3$ that are rigorously known to be strictly jammed \cite{Torquato_2003_Breakdown} under periodic or hard-wall boundary conditions, and hyperuniform.  In such situations, randomly removing a finite fraction of particles such that  there are no ``di-vacancies'' in the two-dimensional example and no ``tri-vacancies'' in the three-dimensional example \cite{Torquato_2003_LatticeBased} while maintaining strict jamming results in non-hyperuniform packings, i.e., $S(0) \neq 0$. This example illustrates vividly that hyperuniformity is degraded by ``defects''---an issue that we will discuss in more detail in the Conclusions. Therefore, the conjecture includes the saturation condition.  Moreover, we know that collisions in equilibrium hard-sphere configurations on the approach to jammed ordered states are not strictly hyperuniform due to vibrational fluctuations and only become exactly hyperuniform when the ideal jammed state without any defects is attained.  We expect this to be the case on the approach to disordered jammed states.  Thus, based on these considerations, one expects that statistically homogeneous disordered strictly jammed saturated packings of identical spheres are hyperuniform.

Importantly, the conjecture eliminates packings that may have a rigid backbone but possess ``rattlers'' (particles that not locally jammed but are free to move about a confining cage) because a strictly jammed packing cannot contain rattlers \cite{Donev_2005_PCF,Torquato_2010}. Typical packing protocols that have generated disordered jammed packings tend to contain a small concentration of rattlers; because of these particles, one cannot say that the whole (saturated) packing is ``jammed''.  Therefore, the conjecture cannot apply to these packings---a subtle point that has not been fully appreciated. 
Nonetheless, it is an open question what effect the rattlers have on hyperuniformity in the context of strict jamming and whether there exists a maximally random jammed (MRJ) state with no rattlers in the infinite-volume limit that is exactly hyperuniform.

Donev et al.\ \cite{Donev_2005_Unexpected} set out to see to what extent relatively large MRJ-like sphere packings (with system sizes of up to $N=10^6$ particles) in $\reals^3$ were hyperuniform, even though there was a small concentration of rattlers (about 2.5\%), precluding them from the conjecture as noted above.  Nonetheless, they found that a packing of $10^6$ particles
that included the rattlers was nearly hyperuniform with $\lim_{k \rightarrow 0} S(k) = 6.1 \times 10^{-4}$ and first peak value $S_{max} = 4.1$ \cite{foot:effective}.
(When the rattlers were removed, the structure factor at the origin had a substantially larger value, showing that the backbone alone is far from hyperuniform.)
This numerical finding supporting the link between effective hyperuniformity of an isostatic disordered packing to its mechanical rigidity
spurred a number of subsequent numerical and experimental investigations that reached similar conclusions \cite{Zachary_2011_PRL,Zachary_2011_PRE1,Zachary_2011_PRE2,Berthier_2011,Weeks_2011,Hopkins_2012,Dreyfus_2015}. 
In all cases, effective or near hyperuniformity is conferred because the majority of the particles are contained in the strictly-jammed backbone and there are few rattlers.
Indeed, it has been systematically shown that as a hard-sphere system, substantially away from a jammed state, is driven toward strict jamming through densification, $S(0)$ monotonically decreases until effective hyperuniformity is achieved at the putative MRJ state.  Specifically, $S(0)$ was found to approach zero approximately linearly as a function of density from 93\% to 99\% of jamming density, where extrapolating the linear trend in $S(0)$ to jamming density yielded $S(0) = -1 \times 10^{-4}$ \cite{Hopkins_2012}.  This study clearly establishes a correlation between distance to jamming and hyperuniformity, and additionally introduces a ``nonequilibrium index'' describing the interplay between hyperuniformity and a dynamic measure of distance to jamming \cite{foot:NonequilibriumIndex}.

In $\reals^2$, disordered, MRJ-like packings of equal-sized disks are very hard to observe, and it has only recently been shown that highly-disordered, isostatic jammed states exist at all \cite{Atkinson_2014}.
Therefore, it is common to introduce  a size dispersity in order to induce geometrical frustration and increase the degree of disorder in the resulting packings.
However, examining the point configurations derived from the centers of such polydisperse packings could lead one to incorrectly conclude that the packings were not hyperuniform.
Zachary et al.\ demonstrated \cite{Zachary_2011_PRE1} that the proper means of investigating hyperuniformity in this case is through a packing's spectral density $\tilde \chi (k)$; that is, making an extrapolation towards the origin to estimate $\lim_{k \rightarrow 0} \tilde \chi(k) = 0$.
They found that MRJ-like binary systems of disks with size ratio $\alpha = 1.4$ and small disk mole fraction $x=0.75$ exhibited near hyperuniform behavior with $\lim_{k \rightarrow 0} \tilde \chi(k) = 1.0 \times 10^{-5}$.
Thus, even though polydispersity is not part of the original conjecture \cite{Torquato_2003_Hyperuniform}, effective hyperuniformity can be observed in polydisperse packings as well, provided that the size distribution is suitably constrained.
It is even possible that  the conjecture can be extended to polydisperse strictly jammed saturated packings; however, necessary and sufficient conditions for this criterion are highly nontrivial.
Nonetheless, jamming is again a crucial necessary property to attain near-hyperuniformity in $\reals^2$ as it was in $\reals^3$.

A fascinating open question remains as to whether putative MRJ packings can be made to be even more hyperuniform than established to date or exactly hyperuniform with numerical protocols as the system size is made large enough.
This is an extremely delicate question to answer because one must be able to ensure that true jamming is achieved to within a controlled tolerance as the system size increases  without bound.
The latter condition is required to ascertain the infinite-wavelength  hyperuniformity property and yet any packing algorithm necessarily must treat a finite system and hence the smallest accessible positive wavenumber at which $S(k)$ or ${\tilde \chi}(k)$ can be measured is of the order of $2\pi/N^{1/d}$, where $N$ is the number of particles.
The situation is further complicated by noise at the smallest wavenumbers, numerical and protocol-dependent errors, and the reliance on extrapolations of such uncertain data to the zero-wavenumber limit.
To  make matters even more complex, we will present evidence  that current packing algorithms stop short of hyperuniformity---and jamming---because requisite collective rearrangements of the particles become practically impossible as criticality is approached, i.e., a type of ``critical slowing down'' \cite{Barber_1983,Binney_1992}. 

\begin{figure}[bthp]
  \begin{centering}
    $
    \begin{array}{cc}
    \includegraphics[width=0.22\textwidth,clip]{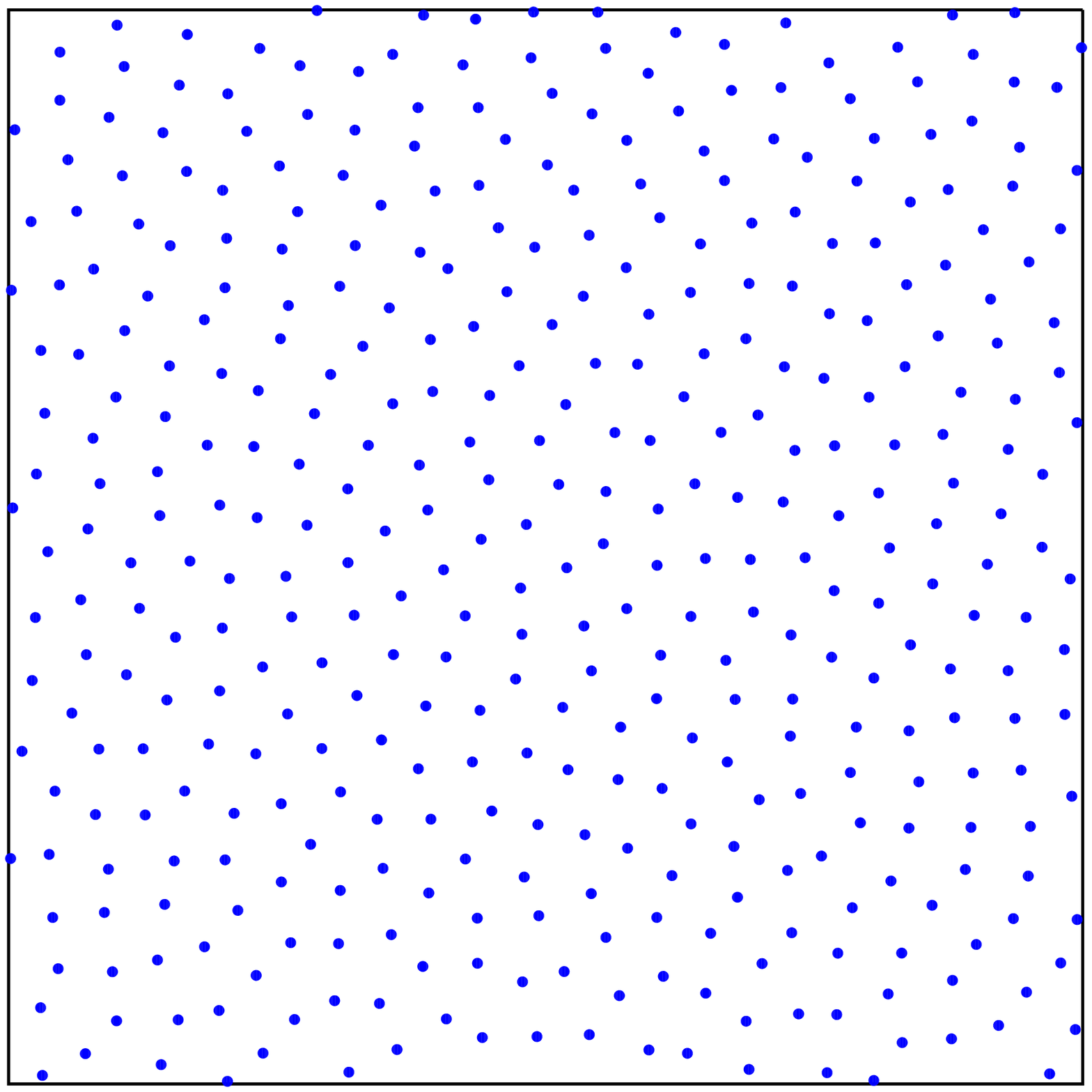}
    &
    \includegraphics[width=0.22\textwidth,clip]{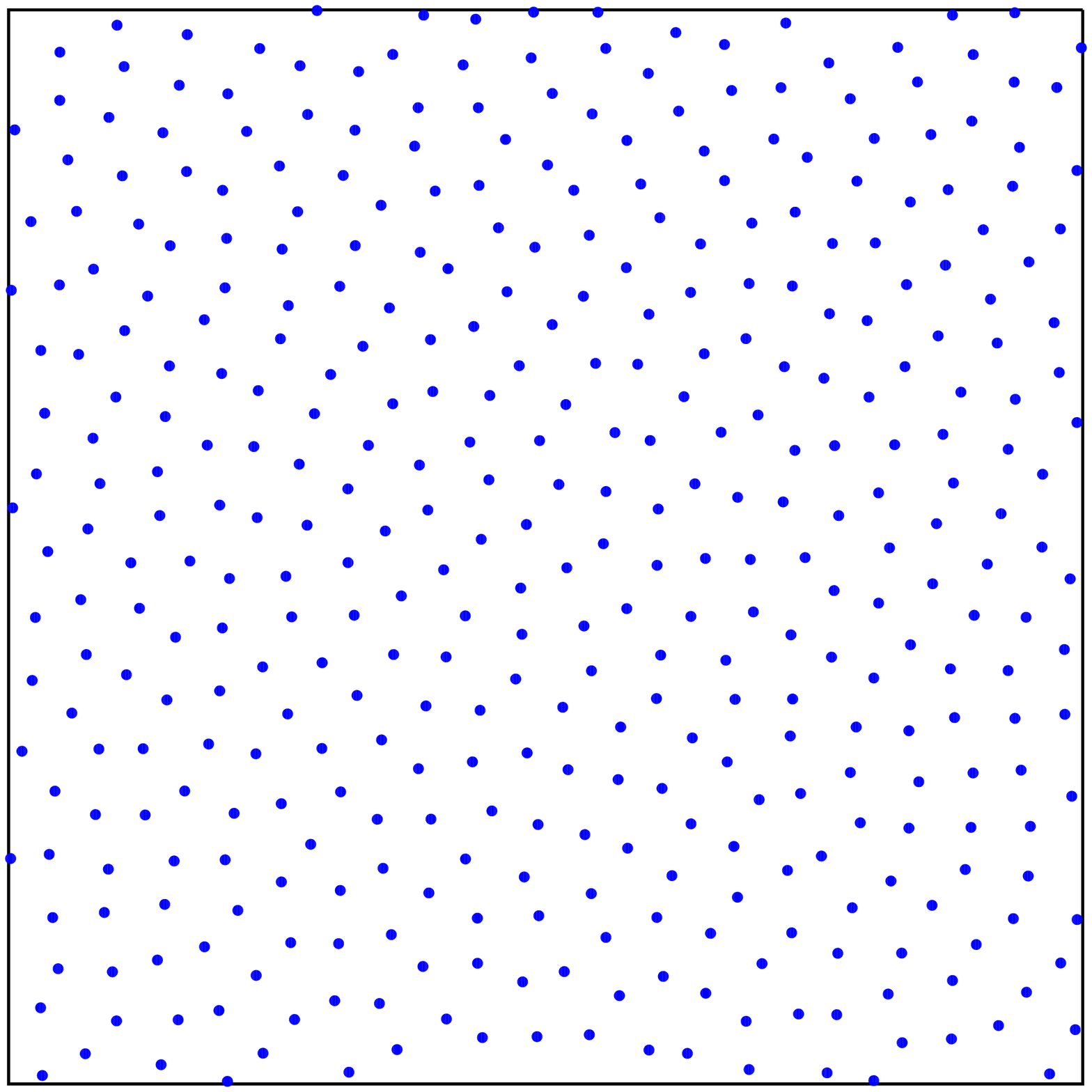}
    \\
    \mbox{(a)}
    &
    \mbox{(b)}
    \end{array}
    $
    \caption{(Color online.) A disordered {\it nonhyperuniform} configuration with $S(0)=1 \times 10^{-4}$ (left panel)
and a disordered {\it hyperuniform} configuration in which the structure factor $S(k)$ vanishes linearly
with $k$ in the limit $k \rightarrow 0$ (right panel). The configuration on the right is obtained by very small collective displacements of the particles on the left
via the collective-coordinate methods described in Ref. \cite{Uche_2006}. Visually, these configurations look very similar to one another, vividly
revealing the ``hidden order'' that can characterize disordered hyperuniform systems \cite{Torquato_2015}.
Indeed, each particle in the left panel on average moves a root-mean-square distance that is about 0.2\% of the 
the mean-nearest-neighbor distance as measured by the configuration proximity metric \cite{Batten_2011} to produce the configuration
in the right panel.
}
\label{fig:CollCoordExample}
\end{centering}
\end{figure}

In this regard, it is noteworthy that general nearly hyperuniform point configurations  can be made to be exactly hyperuniform by very tiny collective displacements via the collective-coordinate approach \cite{Uche_2006}, which by construction enables the structure factor to be constrained to take exact targeted values at a range of wavevectors, as shown recently in Ref.\ \cite{Torquato_2015}.
Figure \ref{fig:CollCoordExample} vividly illustrates this point using an initial nearly hyperuniform configuration in which   $S(0)=1 \times 10^{-4}$ (which is comparable to the value obtained in MRJ-like states) and then collectively displacing the particles by tiny amounts until the structure factor $S(k)$ vanishes linearly with $k$ in the limit $k \rightarrow 0$, as in the case of disordered jammed packings.
While these particles are not jammed, this example serves to emphasize that it only requires very tiny displacements to make a nearly hyperuniform system exactly hyperuniform.
Thus, a critical slowing down implies that it becomes increasingly difficult numerically to drive the value of $S(0)$ down to its lowest possible value if a true jammed critical state could  be attained.

In this paper, we will investigate the role of {\it jamming} and present evidence that suggests that deviations from hyperuniformity 
in MRJ-like packings can in part be explained by a shortcoming of the numerical protocols to generate exactly-jammed configurations as a result of the type of critical slowing down mentioned above.  In order to attempt to observe jammed states, we will utilize a variety of standard hard-sphere packing protocols including the Lubachevsky-Stillinger (LS) event-driven molecular dynamics algorithm \cite{Stillinger_1990,Stillinger_1991} and the Torquato-Jiao (TJ) sequential linear programming algorithm \cite{TJ_2010} to obtain putatively collectively jammed and strictly jammed MRJ packings \cite{foot:MRJ}, 
respectively. 

We will focus on frictionless binary disk packings in two dimensions because it allows us to study the behavior at smaller wavenumbers than three dimensions, assuming $N$ is held constant.  In addition, it is computationally easier to ensure proper jamming at a given system size, further increasing our ability to query the long-wavelength-behavior of the MRJ state.
Studying such systems will enable us to make contact with the recent investigation of Wu et al.\ \cite{Teitel_2015} who examined
binary packings of soft-disks  above the jamming transition. They found that  at finite positive pressures the spectral density exhibited a local minimum at a finite wavenumber and proceeded to grow for smaller wavenumbers, calling into question whether hyperuniformity is observed when approaching the jamming transition from above. 
However, they also  recognized that the presence of hyperuniformity at jamming may be sensitive to the specific protocol used to construct the jammed configurations. 

The rest of the paper is organized as follows: in Sec.\ \ref{sec:Jamming}, we present a variety of methods to test whether a packing, ordered or not, 
is truly jammed.  These tests include a modification of the rigorous linear programming method of Donev et al.\ \cite{Donev_2004_LP}
and so-called ``pressure-leak'' tests.  We then apply them to our packings and find that standard protocols fail to produce jammed packings at surprisingly low system sizes.  
In Sec.\ \ref{sec:Hyperuniformity}, we investigate the subtleties related to system size and numerical protocol that affect one's ability to observe hyperuniform configurations in putatively jammed disordered packings using computer simulations. Discussion and concluding remarks are given in Sec.\ \ref{sec:Conclusion}.

\section{Methods to test for jamming}
\label{sec:Jamming}
A variety of methods have been used to test for jamming in hard-particle packings in the past; we will begin by reviewing some common methods, then introduce the linear programming method that we use in this work to rigorously test packings for jamming.  We will show that even relatively small 2D packings with a high reduced pressure $P = pV/(Nk_BT)$ may not be jammed, even though some methods may imply otherwise.

\subsection{Pressure-Leak Method}

An effective heuristic means of testing for collective jamming is a so-called ``pressure leak'' test \cite{Stillinger_1990,Stillinger_1991,Torquato_2001,Donev_2004_LP}, in which the spheres are subjected to standard molecular dynamics for some relatively large time \cite{foot:PopStrict}. 
If the system pressure begins to drop substantially, then one may conclude that the packing was not collectively jammed; the pressure leak indicates that the particles have discovered an unjamming motion.  While this test is effective for packings that are not well-jammed or have a large interparticle gap, it struggles with packings that are at a high reduced pressure and packings that are nearly jammed, but for which an unjamming motion requires the cooperative motion of many spheres.  In configuration space, this scenario is analogous to the $Nd$-dimensional configuration point being locally confined to a high-dimensional ``bottleneck'' from which escape may only occur in very specific directions.  In such cases, the algorithm may require to process prohibitively many collisions per particle.  Animations demonstrating the pressure test on an ordered and a disordered configuration are included in the Supplemental Material \cite{SupplementalMaterial}.

\subsection{Linear Programming-Based Approach}
Donev et al.\ introduced a method that uses randomized sequential linear programming to rigorously test for collective or strict jamming in packings of frictionless spheres \cite{Donev_2004_LP,Donev_2004_Jamming}.  Speaking physically, the algorithm applies random body forces to the spheres in the packing and seeks to maximize the work due to those forces by displacing them in the direction of their applied forces while obeying the constraint that no spheres overlap.  Spheres that displace as a result of the optimization can be identified as rattlers; if every sphere is a rattler, then the packing is unjammed.

This is implemented using sequential linear programming techniques.  Let $\vc R = (\vc r_1 , \dots , \vc r_N)^T$ describe the position of the $N$ spheres at the beginning of an iteration, and let $\Delta \vc R = (\Delta \vc r_1 , \dots , \Delta \vc r_N)^T$ be a vector of design variables describing how the spheres displace.  The vector $\vc B \in \reals^{Nd}$ contains the body forces which will attempt to displace the spheres; the scalar objective function $Z=\vc B^T \Delta \vc R$ to be maximized is, physically speaking, the work performed on the packing due to $\vc B$.

For two spheres $i$ and $j$ not to overlap, we require $\norm{(\vc r_i + \Delta \vc r_i) - (\vc r_j + \Delta \vc r_j)} \ge D_{ij}$, where $D_{ij} = (D_i+D_j)/2$ is the additive diameter between spheres $i$ and $j$ with diameters $D_i$ and $D_j$.  Linearizing this gives the following linear program (LP):

\begin{eqnarray}
&&{\rm maximize}~ Z=\vc B^T \Delta \vc R
\nonumber
\\
&&{\rm subject~to}
\nonumber
\\
&&\Delta \vc r_i - \Delta \vc r_j \le r_{ij}-D_{ij} ~ \forall~i,j \not = i
\label{eqn:LPJam}
\end{eqnarray}
where $r_{ij} = \norm{\vc r_i - \vc r_j}$.  In the case where one wishes to check for strict jamming, one adds $d(d+1)/2$ strain variables to deform the fundamental cell and a constraint that the fundamental cell volume $V$ does not increase.  For a packing with periodic boundary conditions (as we consider throughout the current work), these variables enter the constraints through pairs of spheres interacting through periodic boundary conditions; in the case where one is considering a packing with hard walls, these variables would show up in constraint terms involving the boundary.

When dealing with non-ideal packings (i.e. packings thought to be very close to exact jamming but with $\phi_c-\phi>0$ where $\phi_c$ is the jamming density), one must provide for the fact that even backbone spheres will be able to move by some small amount.  Therefore, one is forced to relax the criterion that any sphere that moves must be a rattler.  This was done by introducing a tolerance, i.e., any sphere that moves more than $\Delta_{tol}$ is a rattler.  In addition, one must now ask not whether the packing is exactly jammed or not, but whether or not it is confined to a {\it jamming basin}.   Given a jammed configuration $\vc R_J$ at packing fraction $\phi_c$, a jamming basin $\mathcal J(\vc R_J,\phi_c)$ is defined as the set of points in configuration space for which the {\it only} accessible local packing fraction maximum under continuous displacements corresponds to $\vc R_J$ (modulo rattlers).  The density $\phi^*<\phi_c$ is defined as the highest density at which $\vc R$ may be continuously displaced to arrive at at least one other local maximum; the quantity $\phi_c-\phi^*$ is the ``depth'' of the jamming basin.

It can be difficult in practice to pick a value for $\Delta_{tol}$, or to answer the ``jamming basin'' question definitively.  This is because the available configuration space to a packing is, in general, very complicated, and the impression one obtains of it through this algorithm is dependent on the particular choice of $\vc B$.  Traditionally, $\vc B$ is generated randomly, and the LP is solved iteratively several times in order to begin exploring in the direction of $\vc B$; thus, one might obtain a sense of the distance over which spheres in the packing might displace despite being in a very closely-packed configuration \cite{Donev_2004_Jamming}.  In addition, the variations in the geometry of various jamming basins (even those corresponding to an ensemble of similar packings) are considerable.  Between these two factors, it is very difficult in practice to answer the binary question of whether or not a packing is truly within a jamming basin using the standard LP jamming test.

\begin{figure}[bthp]
  \begin{centering}
    \includegraphics[width=0.4\textwidth,clip]{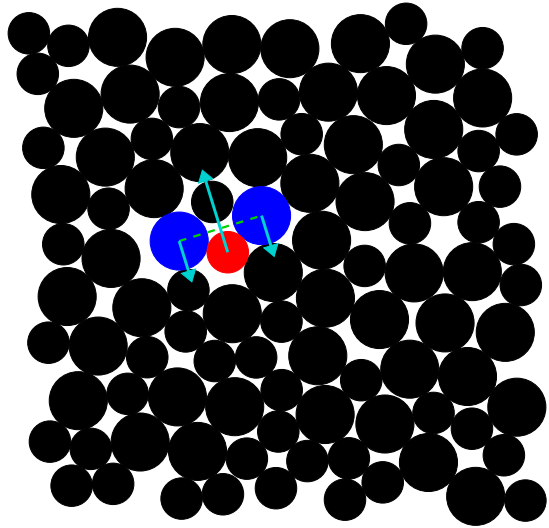}
    \caption{(Color online.) Illustration of the body forces (arrows) chosen for a single iteration of the ``pop test''.  The green dashed line connects the two blue (dark gray when viewed in black and white) spheres that make up $\mathcal C$.  The forces are chosen such that the red (light gray) sphere ``pops'' through the two blue spheres; the sum of the forces on the packing is zero so that uniform translations of the entire packing are not favored.  If the center of the red sphere crosses the green dashed line, then the packing is not confined to a jamming basin.}
    \label{fig:PopTest}
  \end{centering}
\end{figure}

We overcome this difficulty by choosing $\vc B$ in a special manner designed to elucidate a particular local rearrangement that we call a ``pop''.  We pick a backbone sphere $i$ and a $d$-combination of backbone spheres contacting it $\mathcal C=\{j_1,\dots,j_d\}$.  We then determine the plane containing the spheres in $\mathcal C$ and $\vc b$, the unit vector orthogonal to this plane facing away from sphere $i$.  The load applied to the packing is
\begin{equation}
  \vc B = 
  \begin{cases} 
    \vc b    & {\rm for~sphere~}i\\
    -\vc b/d & {\rm for~spheres~in~} \mathcal C\\
    0        & {\rm otherwise.}
  \end{cases}
\nonumber
\end{equation}
This is illustrated in Fig.\ \ref{fig:PopTest}.  Physically, we are asking for sphere $i$ to ``pop'' through the plane described by $\mathcal C$, thus leaving the current jamming basin and entering a new one with a distinct contact network.  It is important that the body forces sum to zero so that trivial uniform translations of the packing are not favored; such movements can obscure whether progress is made in realizing a ``pop''.  Animations demonstrating the pop test on an ordered and a disordered configuration are included in the Supplemental Material \cite{SupplementalMaterial}.

It is important to note that while $\vc B$ is nonzero for only a small number of spheres, the linear program is free to displace {\it all} of the spheres in the packing as it carries out its optimization, meaning that {\it global} rearrangements are being considered.  In other words, while the salient characteristic of a ``pop'' might be a local rearrangement, very complicated collective movements of many particles may take place in effecting it.

We solve the linear program in Eq. (\ref{eqn:LPJam}) iteratively until either sphere $i$ passes ``pops'' through the plane or the packing stops rearranging.  The latter may be inferred by monitoring the objective value associated with an optimization iteration.  In the former case, we have found an unjamming motion, and the packing is unjammed; this may be confirmed by compressing the packing from this new configuration and comparing the ensuing contact network to the original network.  If a pop is not found, we continue by picking another combination of contacts or another sphere until all possibilities are exhausted, at which point we conclude that the packing must be jammed.  By focusing on particle displacements instead of density increases, this test ``inverts'' the jamming problem and is able to efficiently determine if a given packing is within a jamming basin, even when improving the packing density is computationally difficult or the ``pressure test'' fails to find an unjamming motion even after considerable computational time.

%
%

\section{Comparison of Jamming Tests}
\label{sec:JammingResults}

In order to establish a benchmark for the pressure test's performance, we investigated the case of the square lattice and found that when the interparticle gap is on the order of $10^{-12}$ disk diameters, evidence of a developing pressure leak might only become visible after up to $10^6$ collisions per sphere; see Appendix \ref{apx:SquareLattice} for details.

\begin{figure}[bthp]
  \begin{centering}
    \includegraphics[width=0.4\textwidth]{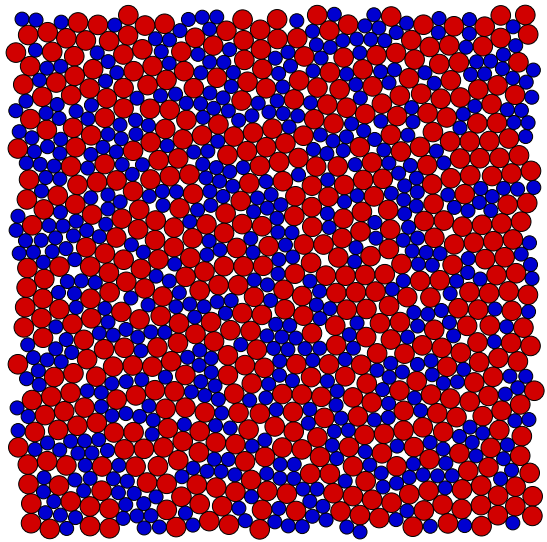}
    \caption{(Color online.) A collectively jammed packing produced by the LS algorithm under the compression schedule specified in Sec.\ \ref{sec:Jamming}.}
    \label{fig:PrettyPacking}
  \end{centering}
\end{figure}

In order to illustrate the difference between the ``pressure test'' and ``pop test'', we consider putatively collectively jammed disordered bidisperse disk packings made using a standard compression schedule with the Lubachevsky-Stillinger (LS) algorithm within a square box with periodic boundary conditions.  The LS algorithm is capable of creating packings that are collectively jammed, but {\it only} if one is careful to design a compression schedule that is slow enough to allow the packing to escape unstable mechanical equilibria by discovering the requisite collective particle rearrangements.  On the other hand, compression rates that are too slow allow the system to equilibrate towards less disordered states, which is at odds with our desire to probe the MRJ state \cite{Truskett_2000}.  To accomplish this, we start from initial conditions produced by random sequential addition at a density of $\phi_0 = 0.40$.  We use an initial dimensionless expansion rate of $\gamma = dD_{max}/dt \sqrt{m/(k_BT)} = 10^{-3}$ where $D_{max}$ is the diameter of the largest disk and $m$ is the mass of a disk and proceed until the reduced pressure exceeds $P=10^6$; at that point, the expansion rate is reduced to $\gamma= 10^{-6}$ and the procedure continues until the reduced pressure reaches at least $P=10^8$.  We considered system sizes of $N=10^2~{\rm to}~10^4$, and we produce between $10^2$ and $10^3$ packings for each system size as is needed for reliable statistics.  In order to encourage the protocol to generate disordered configurations, we consider binary packings with a size ratio $a=1.4$ and number ratio $x=0.5$.  An example is shown in Fig.\ \ref{fig:PrettyPacking}.  At $N=10000$, the mean density of our packings is $\overline \phi = 0.8474$ with a standard deviation of $\sigma(\phi) = 1.8 \times 10^{-4}$ \cite{Tian_2015}.

To provide a basis for comparison, we also used the Torquato-Jiao (TJ) sequential linear programming method to generate collectively jammed MRJ packings with the same size ratio, number ratio, and system size.  Details of the algorithm, which solves the ``adaptive shrinking-cell'' optimization problem, can be found in Ref.\ \cite{TJ_2010}.  For our current work, we use an influence sphere of radius $\gamma_{ij}=D_{max}/10$, translation limit $\left| \Delta x \right| \le D_{max}/20$, and global strain limit $\left| \epsilon_{ij} \right| \le D_{max}/20$.  Initial conditions are created using random sequential addition at a density of $\phi = 0.40$ inside a square fundamental cell of unit volume with periodic boundary conditions.  The packings are compressed while holding the box shape fixed until the volume changes by less than $10^{-12}$ over two successive iterations.  In order to control for distance to jamming, the  density of the terminal configurations are decreased so that their short-time reduced pressure (measured over 1000 collisions per disk) is $P=10^9$.

\begin{figure}[bthp]
  \begin{centering}
    \includegraphics[width=0.4\textwidth,clip]{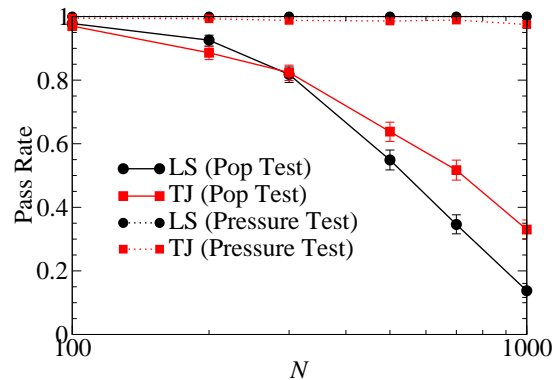}
    \caption{(Color online.) The probability that a packing of a given system size $N$, created using the LS and TJ algorithms, will be jammed, as determined by the ``pop'' and ``pressure'' tests.  Error bars correspond to a 95\% confidence interval as calculated using the Clopper-Pearson method \cite{Clopper_1934}.}
    \label{fig:JamSuccessRate}
  \end{centering}
\end{figure}

We subject our packings to both a pressure test lasting for $10^6$ collisions per disk as well as the ``pop'' test outlined above.  Importantly, {\it nearly all of the LS and TJ packings that we produced pass the pressure test}.  Figure \ref{fig:JamSuccessRate} shows the probability that a packing generated using this compression schedule passes the pop test.  A curve is also included for the TJ algorithm's pass rate for the pressure test.  Our results are consistent with the intuition that jammed packings are more difficult to produce as $N$ increases.  However, our results illustrate vividly that standard methods to assess jamming can give misleading results even for modest system sizes, i.e., as the system size becomes on the order of one thousand disks.

Our results show that one must be careful in assuming that a given protocol is producing packings that are [nearly] truly jammed.  Observing a high reduced pressure that persists for an extended period of time is {\it not} a reliable means of demonstrating that a packing is jammed.  The test we demonstrate here is an efficient means of determining collective motions that unjam packings, revealing that even packings of modest size which were previously thought to be jammed may not actually be so.  As the system size increases, this becomes an increasingly subtle, yet crucial point to which traditional methods like the pressure test are not sensitive.  Given this, one must be careful when relying on numerical results when drawing conclusions about the nature of the MRJ state.  Our results also suggest that previous studies of large, disordered, putatively jammed packings were not carried out on truly-jammed configurations.

\section{Considerations That Prevent Numerical Packings From Being Exactly Hyperuniform}
\label{sec:Hyperuniformity}

In the following subsections, we will examine the effect of system size as well as the packing protocol used when measuring hyperuniformity in nearly-jammed, finite packings of disks.  To do this, we quantify density fluctuations in real and reciprocal space using the local volume fraction variance $\sigma_\tau^2(R)$ and isotropic spectral density $\tilde \chi (k)$, respectively.

In packings of equal-sized spheres, one may investigate the presence of hyperuniformity by considering the sphere centers and computing either their local number density variance $\sigma^2(R)$ or structure factor $S(k)$.  However, these approaches fail to take into account the effect of polydispersity and have been shown \cite{Zachary_2011_PRL,Zachary_2011_PRE1,Zachary_2011_PRE2} to incorrectly suggest that MRJ packings of polydisperse or anisotropic particles are not hyperuniform, whereas $\sigma_\tau^2(R)$ and $\tilde \chi(k)$, which properly account for these particle characteristics, show otherwise.

We begin by reviewing the procedure for computing the spectral density $\tilde \chi(\vc k)$ \cite{Torquato_2002} and local volume fraction variance $\sigma_\tau^2(R)$ \cite{Lu_1990,Zachary_2009} for a packing of polydisperse spheres in $\reals^d$.  For complete derivations, see Ref.\ \cite{Zachary_2011_PRE1}.  For a system of hard-spheres with periodic boundary conditions, the spectral density may be defined via discrete Fourier transform as
\begin{equation}
\tilde \chi (\vc k) = \frac{\left| \sum_{j=1}^N \exp (-i \vc k \cdot \vc r_j) \tilde m(\vc k;D_j/2) \right|^2}{V} ~ (\vc k \not = \vc 0),
\label{eqn:SD_DFT}
\end{equation}
where
\begin{equation}
\tilde m (\vc k;R) = 
\left( \frac{2 \pi}{k R} \right)^{d/2} R^d J_{d/2}(kR)
\label{eqn:FTIndicator}
\end{equation}
is the Fourier transform of the indicator function for a $d$-dimensional sphere of radius $R$ \cite{Torquato_2003_Hyperuniform}; $J_\nu(x)$ is the Bessel function of the first kind of order $\nu$.  The vectors $\vc k$ at which this may be evaluated are integer combinations of the reciprocal basis vectors, defined as the columns of the matrix $\mtx \Lambda_R = [(2\pi) \mtx \Lambda^{-1}]^T$, where the columns of $\mtx \Lambda \in \reals^{d \times d}$ span the fundamental cell of our simulation box.

In order to investigate the nature of density fluctuations in real space, one may consider the variance of the local volume fraction, defined as \cite{Lu_1990}
\begin{equation}
\sigma_\tau^2(R) = \frac{1}{v_1(R)} \int_{\reals^d} \chi(\vc r) \alpha(r;R) d \vc r,
\label{eqn:VolFracVariance}
\end{equation}
where $v_1(R)$ is the volume of a $d$-dimensional sphere of radius $R$, $\chi(\vc r)$ is the autocovariance function, and $\alpha(r;R)$ is the {\it scaled intersection volume}, which is as the intersection volume of two spheres of radius $R$ separated by a distance $r$ divided by $v_1(R)$.  In practice, $\sigma_\tau^2$ may be computed by randomly placing a sufficiently large number of spherical windows of radius $R$ within the packing.

\subsection{Hyperuniformity and System Size}

\begin{figure}[bthp]
  \begin{centering}
    \includegraphics[width=0.4\textwidth,clip]{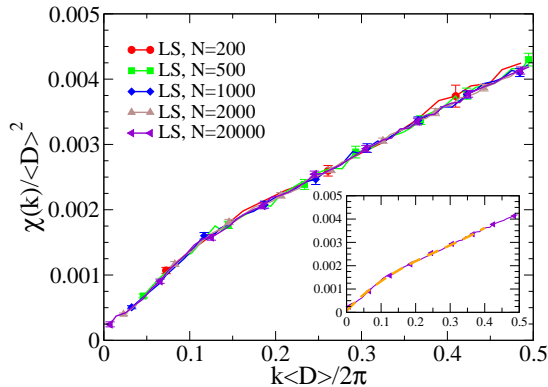}
    \caption{(Color online.) Spectral density for LS packings with various system sizes.  Curves shown are binned ensemble averages with 1000 packings per curve and bin width $\Delta k=0.01$.  While ensembles of smaller system sizes seem to be hyperuniform, a saturation in $\tilde \chi(k)$ is observed as the system size increases and jamming is no longer ensured.  Error bars are shown for a 95\% confidence interval.  The inset shows the cubic fit reported in Table \ref{tab:Chi0} (orange dashed line) on top of the binned data for $N=20000$.}
    \label{fig:Spectral_SysSize}
  \end{centering}
\end{figure}

In order to study the relation between jamming and hyperuniformity in disordered packings that are putatively jammed, we use the LS algorithm with the compression schedule described above to produce packings of binary disks with system sizes up to $N=2 \times 10^4$.  In particular, we will establish that not only does jamming become less common as $N$ increases (as shown above), but that hyperuniformity is concomitantly lost to a ``saturation'' in $\tilde \chi(k)$ at small wavenumbers.  The spectral densities of our packings are plotted in Fig.\ \ref{fig:Spectral_SysSize}; the curves drawn are  ensemble averages with 1000 packings per curve; data is binned according to wavenumber $k = \norm{\vc k}$ with a bin width of $\Delta k = 0.01$.
Raw data for the spectral density of the packings in these ensembles is available in the Supplemental Material \cite{SupplementalMaterial}.

There are significant variations within any ensemble from packing to packing; this variation and its effect on determining hyperuniformity will be discussed for a set of 1000 packings of $N = 500$ binary disks produced using the LS protocol in this section.  To quantify this, we fit the unbinned spectral density of each packing individually with a  polynomial of order $n$ (i.e. $f(\left| \vc k \right|;a_0,\dots,a_n) = \sum_{j=0}^n a_j \left| \vc k \right|^j$) for all wavenumbers within $0 < \left| \vc k \right| \langle D \rangle / 2 \pi \le k_{max}$, where $\langle D \rangle = 1/N \sum_{i=1}^N D_i$ is the number-averaged diameter.  For $n=1$ and $2$, we use $k_{max} = 0.11$ to investigate the behavior below the kink seen in Fig.\ \ref{fig:Spectral_SysSize}.  For $n=3$, we use $k_{max}=0.40$ for more complete data.  We pick this particular value of $k_{max}$ for $n=3$ because the standard deviation of the value of the fit's intercept is minimized for this range.  To illustrate the goodness of fit typical of our fits, the inset of Fig.\ \ref{fig:Spectral_SysSize} shows binned data for the $N=20000$ ensemble along with its fitted cubic polynomial.

For our ensemble of $N=500$ fitted with a cubic polynomial, the intercept at the origin $a_0$ has a mean of $6.5 \times 10^{-5}$ and a standard deviation of $1.8 \times 10^{-3}$, meaning that these packings can be considered {\it effectively hyperuniform} in the sense that if one were to consider a single packing from this ensemble and ask whether is is hyperuniform, then the data imply that the random variations from packing to packing are large enough that one could conclude that the answer is ``yes'' within this noise.  Interestingly, the volume fraction fluctuations in direct space strongly corroborate this conclusion, suggesting that it is of considerable utility when diagnosing hyperuniformity in small systems; we will look at this presently.  The mean extrapolated values of $\tilde \chi(0)$ for the other ensembles of packings shown in Fig.\ \ref{fig:Spectral_SysSize} are presented in Table \ref{tab:Chi0} for various $n$ and $k_{max}$.

\begin{table}[bthp]
  \begin{center}
  \caption{Extrapolated spectral densities at $k=0$ for ensembles of $N_p$ packings created with the LS algorithm with system size $N$, fitted using a polynomial of order $n$ for data at wavenumbers $0 < k \langle D \rangle / 2 \pi  \le k_{max}$.  Reported values are the mean $\overline{a_0}$ and standard deviation $\sigma(a_0)$ for ensembles of $1000$ packings for $N \le 2000$ and 100 packings for $N=20000$.}
  \begin{tabular}{|r|r|r|r|r|r|}
    \hline $N$   & $N_p$ & $n$ & $k_{max}$ & $\overline{a_0}$      & $\sigma(a_0)$ \\
    \hline 200   & 1000 & 3 & 0.40 & $-1.2 \times 10^{-4}$ & $4.2 \times 10^{-3}$  \\

    \hline 500   & 1000 & 3 & 0.40 & $ 6.5 \times 10^{-5}$ & $1.8 \times 10^{-3}$  \\

    \hline 1000  & 1000 & 1 & 0.11 & $ 8.6 \times 10^{-5}$ & $6.5 \times 10^{-4}$  \\
                 &      & 2 & 0.11 & $ 1.6 \times 10^{-4}$ & $1.7 \times 10^{-3}$  \\
                 &      & 3 & 0.40 & $ 1.0 \times 10^{-5}$ & $1.1 \times 10^{-3}$  \\

    \hline 2000  & 1000 & 1 & 0.11 & $ 8.3 \times 10^{-5}$ & $4.1 \times 10^{-4}$  \\
                 &      & 2 & 0.11 & $ 6.2 \times 10^{-5}$ & $8.7 \times 10^{-4}$  \\
                 &      & 3 & 0.40 & $ 5.5 \times 10^{-5}$ & $6.8 \times 10^{-4}$  \\

    \hline 20000 &  100 & 1 & 0.11 & $ 8.5 \times 10^{-5}$ & $1.2 \times 10^{-4}$  \\
                 &      & 2 & 0.11 & $ 1.5 \times 10^{-4}$ & $1.9 \times 10^{-4}$  \\
                 &      & 3 & 0.40 & $ 6.5 \times 10^{-5}$ & $2.2 \times 10^{-4}$  \\
    \hline                                                        
  \end{tabular}
  \label{tab:Chi0}
  \end{center}
\end{table}

As the system size grows, a deviation from hyperuniformity becomes increasingly apparent in the spectral density curves, i.e., a ``saturation'' appears at low wavenumbers; the curvature of the trend inflects at about $k\langle D \rangle / (2 \pi) = 0.05$.  Interestingly, this wavenumber roughly corresponds to the largest wavelength that is typically observable in a jammed system of 500 disks.  At the same time, this system size lies approximately at the point where a given packing will be jammed with a likelihood of about 50\%.

This prompts us to suggest that the physical origin of the ``saturating'' behavior may be fundamentally linked to the ability of the packing protocol to resolve collective rearrangements on a corresponding length scale that may be necessary to reach a truly-jammed state.  A point in  configuration space may become locally trapped within a ``bottleneck'' from which escape may only happen in very few directions.  The result in practice is a ``critical slowing down''\cite{Barber_1983} in which exact jamming takes increasingly long to resolve, as we elaborate in Sec.\ \ref{sec:Conclusion}.

We have also found that this ``saturating'' behavior can be misrepresented when care is not taken in extrapolating the effective value of $\tilde \chi (0)$ from the $\tilde \chi (k)$ for which data is available.  Specifically, the effect of binning data to obtain an ensemble average can artificially increase or decrease the perceived value of $\tilde{\chi}(0)$, where we have observed mostly increases in $\tilde{\chi}(0)$ when bin size is substantially larger than the smallest wavenumber.  Moreover, the spectral density of individual packings tend to vary by a significant amount from that of the ensemble average.  This begs the question: how close to zero must the extrapolated spectral density at $k=0$ be in order to be considered ``effectively hyperuniform''?  We address these questions in detail in Appendix \ref{apx:Fitting}.

Our estimates for a single packing in the $N = 2000$ ensemble described in Appendix \ref{apx:Fitting} indicate that $a_0 = \tilde{\chi}(0)$ will be randomly distributed about zero with a standard deviation of about $4 \times 10^{-4}$. Averaging over an entire 100 packing set yields a better estimate of the mean value $\overline{a_0}$ of the average of the set, and tighter zero-hypothesis confidence intervals.  Our methods show that with about 50\% probability, a mean value of $\overline{a_0}$ such that $-4 \times 10^{-5} \leq \overline{a_0} \leq 4 \times 10^{-5}$ indicates effective hyperuniformity, given the noise inherent in the calculation.  Referencing Table \ref{tab:Chi0} values for the $N=2000$ packings and applying Student's t-distribution to the standard deviation reported, we see that with 68\% probability the mean is within $6.1 \times 10^{-5}$ of $4.6 \times 10^{-5}$. This is on the higher side of the range $-4 \times 10^{-5}$ to $4 \times 10^{-5}$, indicating that the packings together could be hyperuniform within the error, but suggesting that individually many of them are not.  This result corresponds well with the prediction that jamming and hyperuniformity are linked, since the majority of the $N=2000$ packings are not jammed according to the pop test, yet they are close to jamming since they all pass the pressure test.

\begin{figure}[bthp]
  \begin{centering}
    \includegraphics[width=0.4\textwidth,clip]{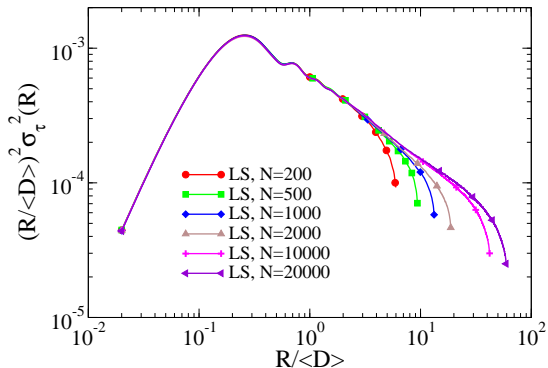}
    \caption{(Color online.)  Window packing fraction variance scaled by the window volume $(R/\langle D \rangle)^2$ for LS packings for various $N$.  Curves shown are binned ensemble averages.  If the ordinate scales towards zero, then the systems are hyperuniform.  The sudden decrease at the highest values of $R$ for each curve are due to finite-size effects, but hyperuniformity is apparent even for the smaller system sizes.  The uncertainty in the curves is very small, on the order of the line width.}
    \label{fig:WindowPF_SysSize}
  \end{centering}
\end{figure}

We also compute the local packing fraction variance $\sigma_\tau^2$ for these ensembles with various system sizes, as shown in Fig.\ \ref{fig:WindowPF_SysSize}.  For hyperuniform packings, $\sigma_\tau^2(R)$ will scale towards zero more quickly than $R^{-2}$ as $R$ tends towards infinity.  Equivalently, $R^2 \sigma_\tau^2(R)$ will be a decreasing quantity with growing $R$.  This is clearly the case with our data, implying that these ensembles are hyperuniform by this metric.  We see a sudden decrease for sufficiently large $R$ for each $N$, but our studies on larger samples imply that this is a finite-size effect similar to that observed in Ref.\ \cite{Dreyfus_2015}.  However, note that the hyperuniformity trends are apparent even for the smaller system sizes.  These results indicate that there is a maximum length scale $R_{max}$ (smaller than the half-width of the simulation box) that can be considered when diagnosing hyperuniformity. In addition, while the spectral density calculation may require hundreds or even thousands of packings to converge to well-resolved curves, the direct-space curves converge much more quickly.  It has been noted before that if one must ascertain hyperuniformity from either a small system or an ensemble with a limited population, the direct-space computation is particularly effective at diagnosing hyperuniformity \cite{Dreyfus_2015}.

\subsection{Hyperuniformity and Protocol Dependence}
\label{sec:HU_Protocols}
We now consider the effect that one's choice of packing protocol has on the degree of hyperuniformity.  In particular, we will present evidence that different protocols, which come near to exact jamming to different degrees, create packings with correspondingly different degrees of hyperuniformity.  Importantly, while all of the protocols considered have the capacity to generate jammed packings given arbitrarily high numerical precision and computing time, practical considerations like computational cost and the rate of convergence force one to terminate any algorithm before exact jamming is attained.  It is this distance to jamming that we would like to focus on.  To do this, we begin with our LS-generated binary disk packings as well as packings made using a standard soft-sphere protocol following the procedure in \cite{Teitel_2015}.  For the latter algorithm, particles interact through the standard harmonic pair potential \cite{Liu_2003}
\begin{equation}
\psi_{ij}(r_{ij}) = 
\begin{cases} 
\frac{1}{2} \left(1 - r_{ij}/D_{ij} \right)^2 & {\rm for}~r_{ij} <   D_{ij} \\
0                   & {\rm for}~r_{ij} \ge D_{ij}
\end{cases}
\end{equation}
where $D_{ij} = (D_i + D_j)/2$ and $D_i$ is the diameter of particle $i$.  An enthalpy-like function $H=\Gamma_N \log(V)+\sum_{i,j} \psi_{ij}$ is minimized for $\Gamma_N/N=1.373 \times 10^{-4}$ using a conjugate gradient method, terminating when the gradient of the objective function falls below $10^{-13}$ or its value remains constant (as expressed in double precision) over 10 optimization steps.  Initial configurations are Poisson point processes at reduced density $\sum_i v_i/V = 0.84$, where $v_i$ is the volume of particle $i$.

We input the resulting packings from these two protocols as initial conditions \cite{foot:SSDiamReduce} 
for the TJ algorithm, and proceed to generate putatively strictly-jammed packings.  An animation demonstrating this combination protocol using the soft-sphere algorithm followed by the TJ algorithm is provided in the Supplemental Material \cite{SupplementalMaterial}.  Since TJ seeks out local packing fraction maxima, {\it the threshold state being approached in the soft-sphere protocol is the same state that is being approached using TJ.}  While it is beyond the scope of this work to investigate the similarities and differences in the approach to the jamming transition from these two protocols, the limiting configuration is essentially unchanged when the packings are given to TJ.

\begin{figure}[bthp]
  \begin{centering}
    \includegraphics[width=0.4\textwidth,clip]{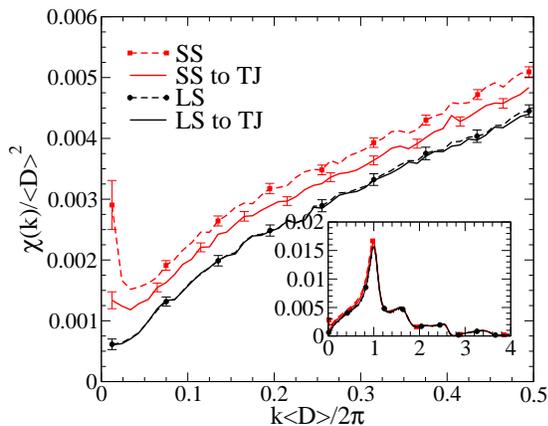}
    \caption{(Color online.) The spectral density of disordered binary disk packings created by a variety of protocols.  Curves shown are binned ensemble averages with bin width $\Delta k=0.01$.  Feeding the packings produced by the soft-sphere protocol (identified as ``SS'' in the legend) into the TJ algorithm increases the hyperuniformity of the packings significantly as they are brought closer to exact jamming.  The putatively collectively-jammed packings generated by LS are almost unchanged upon being given to TJ for strict jamming, implying that the difference between collective and strict jamming is small at large system sizes.  Error bars are shown for a 95\% confidence interval.}
    \label{fig:SD_Protocol}
  \end{centering}
\end{figure}

Figure \ref{fig:SD_Protocol} shows the ensemble-averaged spectral density of ensembles of 100 packings of system size $N=10^4$ for both protocols before and after being input to TJ; the corresponding direct-space measurement of the window volume fraction variance $\sigma_\tau^2(R)$ is shown in Fig.\ \ref{fig:WindowPF}.
Raw data for the spectral density of these ensembles is available in the Supplemental Material \cite{SupplementalMaterial}.
The LS-generated configurations do not change by a significant amount upon being fed to the TJ algorithm, suggesting that the difference between collective and strict jamming is minor at sufficiently large system sizes \cite{Donev_2004_Jamming,Donev_2005_PCF}.

\begin{figure}[bthp]
  \begin{centering}
    \includegraphics[width=0.4\textwidth,clip]{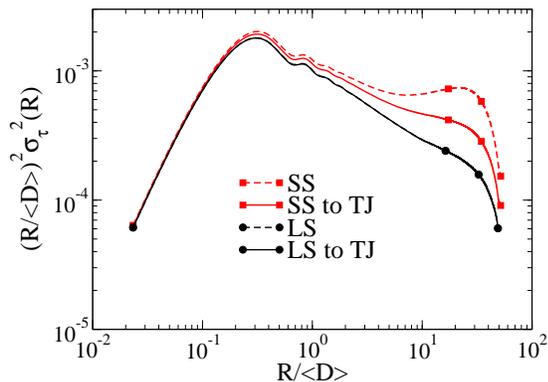}
    \caption{(Color online.) Window number variance scaled by the window volume $(R/\langle D \rangle)^2$ for binary disk packings with system size $N=10^4$ created by a variety of protocols.  Feeding the packings produced by the soft-sphere protocol (identified as ``SS'' in the legend) into the TJ algorithm increases the range over which hyperuniform scaling behavior is observed as the packings are brought closer to exact jamming.  The LS packings exhibit hyperuniform scaling over a significant range of length scales (i.e.\  the ordinate scales towards zero as $R$ increases). The uncertainty in the curves is on the order of the line width.}
    \label{fig:WindowPF}
  \end{centering}
\end{figure}

Of particular importance is that the sharp increase in $\tilde \chi$ near the origin that is observed in soft-sphere packings at a positive pressure \cite{Teitel_2015} vanishes upon subsequent packing using the TJ algorithm.  This is seen in the direct-space calculations as well, since the non-hyperuniform $R^{-2}$ scaling in $\sigma_\tau^2(R)$ seen with the soft-sphere protocol at length scales beyond $R=7 \langle D \rangle$ becomes hyperuniform upon subsequent packing with TJ.  This ability to obtain length scales over which hyperuniformity is observed is one of the main strengths of the direct-space calculation \cite{Dreyfus_2015}.  TJ is allowing us to get several orders of magnitude closer to exact jamming and more accurately discern its spectral density: for TJ, $\phi_J-\phi \approx 10^{-12}$, whereas in the soft-sphere case, $\phi-\phi_J \approx 10^{-3}$.  This reinforces the idea that bringing a packing closer to exact jamming will improve its hyperuniformity.  Additionally, this result again emphasizes the subtleties in characterizing the long-wavelength nature of the MRJ state and underscores the need to be mindful of the nature of the numerical insensitivities in packing protocols---particularly as they approach highly-disordered states.

%
%

\section{Conclusions and Discussion}
\label{sec:Conclusion}

We have introduced a method that uses sequential linear programming to test for jamming in packings of frictionless hard-spheres.  We applied this algorithm to disordered packings of bidisperse disks and found that standard protocols struggle to create packings that are truly confined to a jamming basin when the system size grows to be on the order of one thousand disks or more.  Importantly, heuristic tests like the pressure test fail to find these unjamming motions.  We then examined the spectral density of the packings we generated for a variety of system sizes and found an inflection at a wavenumber corresponding to the largest accessible length scale in a typical packing with system size $N=500$---the same system size for which the probability of producing a truly-jammed configuration is approximately 50\%.  Given this, we conclude that our inability to observe exactly-hyperuniform configurations at larger system sizes is directly linked to the difficulty of producing exactly-jammed configurations.

We also found that, by bringing soft-sphere packings closer to their jamming transition point by using the TJ algorithm, the degree of hyperuniformity was also increased by several orders of magnitude as measured by the volume fraction fluctuations in direct space.  This ability to obtain length scales over which hyperuniformity is observed is one of the main strengths of the direct-space calculation \cite{Dreyfus_2015}.  This finding also suggests that the break from hyperuniformity at large length scales, previously thought to be inherent to packings above the jamming transition, is in reality due to an excessive distance from the jamming transition, and that subsequent resolution makes this feature disappear.  Because of our careful consideration of packings' distance to jamming and the corresponding evolution of their structure, we conclude that one cannot rely solely on current technology to say that there is no connection between jamming in disordered packings and hyperuniformity since jamming cannot be ensured.

In both of the cases studied above, we point out an evident ``critical slowing down'' in that the collective particle movements required to reach an exactly-jammed state take longer to resolve as the system size grows.  A point in configurational space may become locally trapped within a ``bottleneck'' from which escape may only happen in very few directions, corresponding to the collective rearrangements that the packing must undergo.  For an event-driven MD protocol such as LS, it may take many collisions per particle before this escape is discovered.  As the system size increases, the dimensionality of configuration space increases as well and this ``escape'' becomes an increasingly rare event.  The result in practice is a ``critical slowing down'' in which exact jamming takes increasingly long to resolve.  This dynamic critical behavior is well-known in other physical systems, the most well-known of which is perhaps the kinetic Ising model \cite{Barber_1983}.  The fact that it is also observable in packing contributes additional evidence to support the idea that the MRJ state lies at a special type of critical point, namely, one in which the direct correlation function $c(r)$, rather than the total correlation function $h(r)$, is long-ranged due to the fact that the appropriate spectral function is zero at $k=0$ \cite{Torquato_2003_Hyperuniform}.  This is to be contrasted with a thermal critical point in which density fluctuations diverge because $h(r)$ is long-ranged.

It is important to notice that the protocols that we are aware of tend to produce configurations possessing a positive rattler fraction.  As these particles do not contribute to the rigidity of the backbone, they might be regarded as ``defects'' within the disordered configuration.  Their location is also not uniquely specified, in contrast to the positions of the backbone particles.  Therefore, we expect that any packing containing rattlers cannot necessarily be {\it exactly} hyperuniform due to the freedom the rattlers possess.  However, one must also be aware that the backbone configurations that give rise to rattler cages are also interesting in that the cages surrounding the rattlers tend to have significantly different local structures from that of the rest of the packing \cite{Atkinson_2013}.
Therefore, it is not enough to expect that ``optimizing'' the rattlers' positions will necessarily yield a hyperuniform configuration.  In general, one should not expect that a packing possessing a jammed backbone will be hyperuniform unless it is saturated and {\it does not possess any rattlers}.  It has been observed previously \cite{Atkinson_2013} that the TJ algorithm produces packings of equal-sized spheres in three dimensions that are simultaneously more disordered and have significantly fewer rattlers than other known protocols (e.g., \cite{Stillinger_1991}), leaving open the possibility that packings that are more disordered may have even fewer rattlers still.
This raises the possibility that the ideal MRJ (most disordered, strictly jammed) packings of identical spheres have no rattlers. 
This possibility is currently being further investigated \cite{Atkinson_2016_DCF}.
Indeed, if this is true, then one might reasonably expect that this would also hold for two-dimensional systems of identical particles \cite{Atkinson_2014} and certain polydisperse packings in 2D and 3D (given qualifications on the distribution of particle sizes).
Devising algorithms that produce large rattler-free disordered jammed packings is an outstanding, challenging task. 
According to the Torquato-Stillinger conjecture, any MRJ-like strict jammed packing without any rattlers would be hyperuniform in the infinite-volume limit \cite{Torquato_2003_Hyperuniform}.

It has been found that the average rattler fraction $N_R/N$ observed in a disordered packing is dependent upon the protocol being used; in three dimensions, putative MRJ packings produced with the LS algorithm tend to have a rattler fraction of approximately $N_R/N \approx 0.025$ \cite{Donev_2005_Unexpected}.  On the other hand, the TJ algorithm produces packings with $N_R/N \approx 0.015$ \cite{Atkinson_2013}.  For the binary systems considered in this work with $(\alpha,x) = (1.4,0.5)$, the LS algorithm produces an average rattler fraction of $N_R/N=0.063 \pm 0.001$, whereas the TJ algorithm produces a mean of $N_R/N=0.048 \pm 0.001$, mirroring the story in three dimensions.  Given that the packings produced by TJ are significantly more disordered as measured by standard order metrics \cite{Atkinson_2013}, we ask whether the true MRJ state has {\it no} rattlers.  The existence of such a jammed state remains an open question, and addressing it would presumably require a novel packing protocol.



For $d=3$, similar difficulties to those observed in this work exist in producing truly-jammed packings.  For example, previous investigations have suggested that it is difficult to produce disordered packings of $N=10^4$ spheres using the LS algorithm that can pass even a pressure-leak test \cite{Donev_2005_PCF}.  Simulations using TJ take increasingly long amounts of time in producing jammed packings at comparable system sizes.  Thus, we point out that there is an issue that is practical in nature associated with producing jammed packings of monodisperse spheres at large system sizes.  An apparent ``saturation'' was observed in the structure factor of disordered soft-sphere packings in which the trend $S(k) \propto k$ breaks down and becomes constant and positive for smaller wavenumbers, implying that jamming and hyperuniformity may not be connected \cite{Ikeda_2015}.  The wavenumber associated with this ``turnover'' corresponds to wavelengths on the order of approximately 20 spheres, which mirrors the ``critical'' system size above which it seems to be difficult to generate truly-jammed packings.  This parallels the observations we have made in our current study.  We suggest, therefore, that this observed departure from exact hyperuniformity may be due to an inability to resolve the particle displacements necessary to approach the jamming threshold--particularly given the system sizes that were considered ($N=5 \times 10^5$).


Our results demonstrate the particular difficulty of producing systems of either hard or soft particles that fall precisely at the jamming transition.  Moreover, ensuring that a packing is truly in a jamming basin is a highly nontrivial task.  Is there hope for producing improved algorithms to yield higher quality MRJ-like states?  Of critical importance is that the algorithm be efficient at determining and resolving the collective movements that allow the packing to escape configurations that are not true local density maxima.  Evidently, some of these motions may involve all of the particles in the system, and may achieve very small immediate changes in density.  Thus, an intelligent way of identifying and carrying out these displacements is warranted.


We have suggested that reaching an exactly-jammed state requires particles to ``slide" by each other in 
very specific ways when the packing is barely inside a jamming basin.   Nonetheless, we point out that {\it near-hyperuniformity} may be readily observed and quantified in a number of meaningful ways.  One is to measure the ratio between the value of the structure factor at the origin (e.g., as obtained by scattering experiments) and at the first peak \cite{foot:effective}.  A second technique is to measure the volume fraction fluctuations in real space as a function of observation window size and look for the appropriate scaling relation as above \cite{Dreyfus_2015,Hexner_2015}.  This latter technique has indicated the existence of real-world systems that are hyperuniform over length scales spanning as much as four orders of magnitude \cite{Hexner_2015}.  Thus, it is of interest to investigate the physical consequences of near-hyperuniformity as observed in such systems.


\begin{acknowledgments}
The authors thank F.\ H.\ Stillinger for many insightful discussions.  This work was supported in part by the National Science Foundation under Grant No.\ DMS-1211087.
\end{acknowledgments}

\appendix
\section{Jamming tests on the square lattice}
\label{apx:SquareLattice}

\begin{figure}[bthp]
  \begin{centering}
    $
    \begin{array}{cc}
      \includegraphics[width=0.2\textwidth,clip]{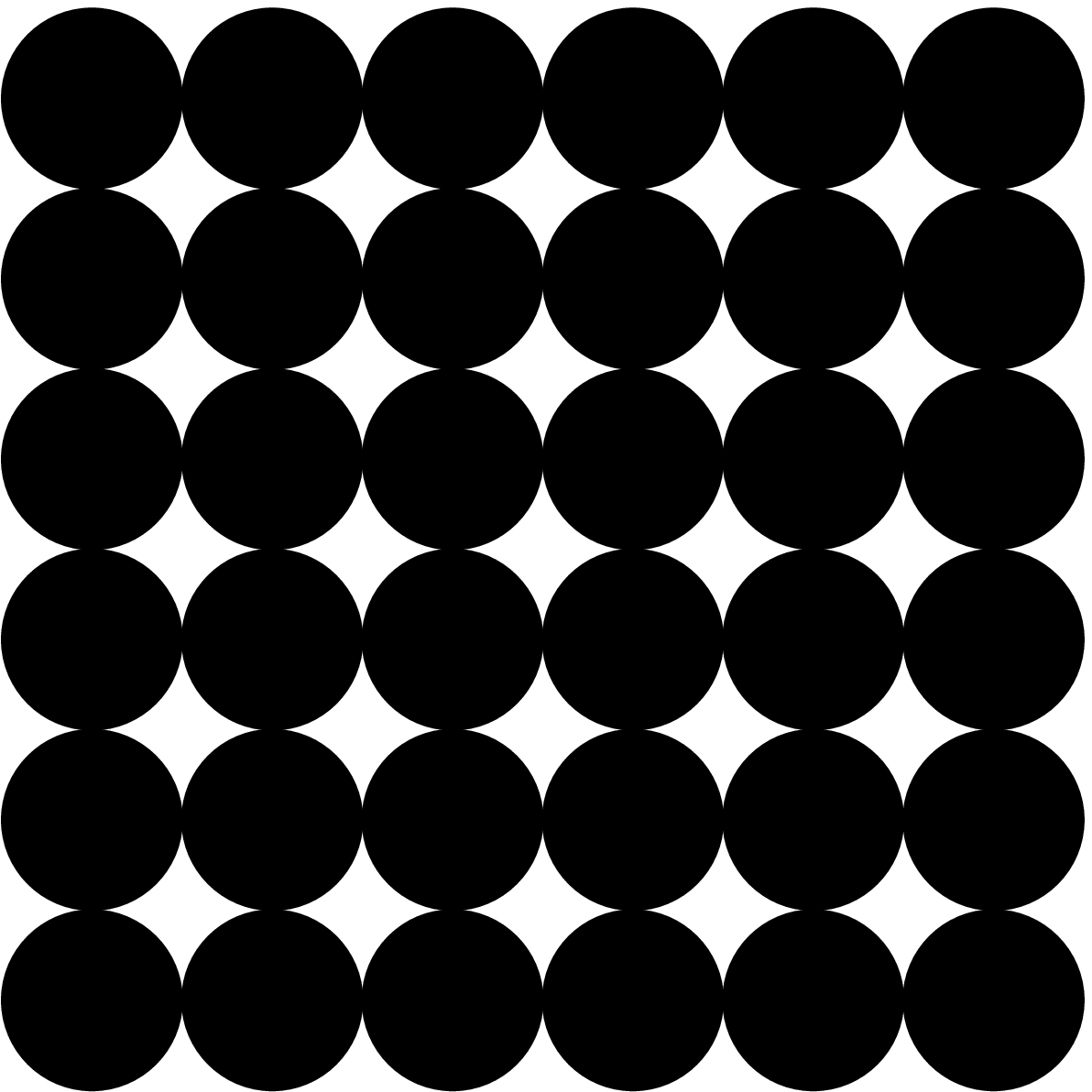}
      &
      \includegraphics[width=0.2\textwidth,clip]{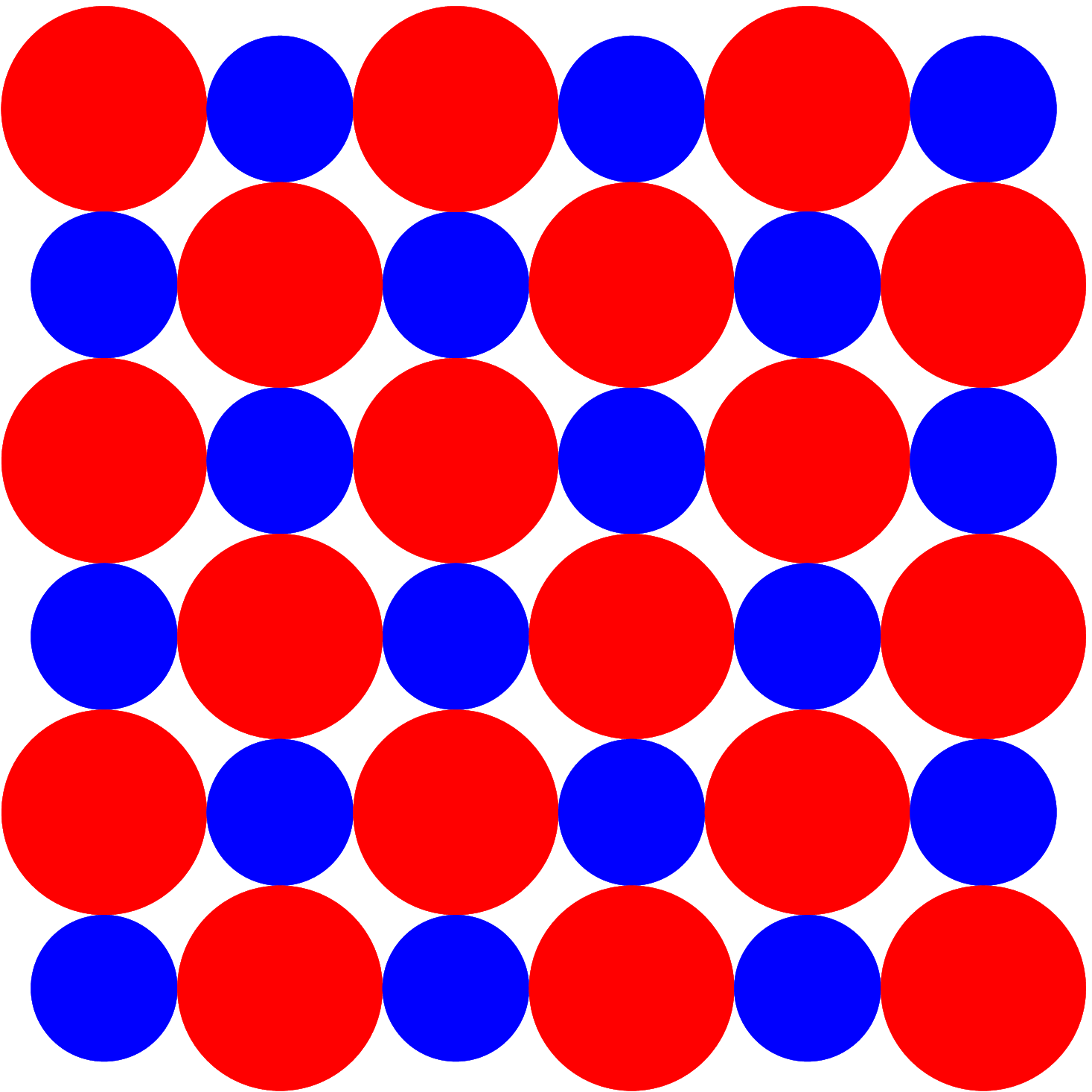}
      \\
      \mbox{(a)}
      &
      \mbox{(b)}
    \end{array}
    $
    \caption{(Color online.) Square lattice packings of (a) monodisperse and (b) binary disks.}
    \label{fig:SquareLatticeICs}
  \end{centering}
\end{figure}

In order to assess the reliability of the pop test for discovering unjamming motions, we consider the case of a square lattice in 2D, using both (a) a monodisperse disk packing and (b) a binary disk packing with size ratio $(\alpha,x)=(1.4,0.5)$; examples of the starting configurations are shown in Fig.\ \ref{fig:SquareLatticeICs}.

\begin{figure}[bthp]
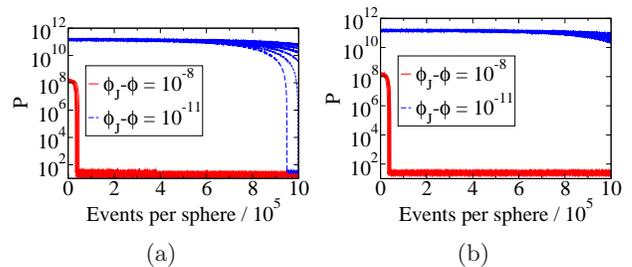

  \begin{centering}
    $
    \begin{array}{cc}
      \includegraphics[width=0.22\textwidth,clip]{./Fig10A.eps}
      &
      \includegraphics[width=0.22\textwidth,clip]{./Fig10B.eps}
      \\
      \mbox{(a)}
      &
      \mbox{(b)}
    \end{array}
    $
    \caption{(Color online.) Pressure tests for a square lattice of (a) monodisperse and (b) binary spheres with $(\alpha,x) = (1.4,0.5)$ with $N=10^2$ disks at putative jamming gaps of $\phi_J-\phi=10^{-8}$ (lower curves in each subfigure) and $10^{-11}$ (upper curves).  As the jamming gap approaches zero, longer simulations are required to observe unjamming.  For systems with $P \approx 10^8$, a simulation of one million collisions per particle is usually long enough to observe an unjamming motion.}
    \label{fig:PressureTestSq}
  \end{centering}
\end{figure}

We considered system sizes of $N=10^2$ and $10^4$ and size ratios $\alpha=1$ and $1.4$ (the former referring to the monodisperse limit).  We perform pressure pressure tests for $\phi_c^*-\phi = 10^{-8}$ and $10^{-11}$, where $\phi_c^*(\alpha) = [ \pi (1+\alpha^2)] / [2(1+\alpha)^2]$ is the close-packing density of the lattice \cite{foot:DensestRange}. 
The reduced pressure is plotted as a function of the number of events per sphere in Fig.\ \ref{fig:PressureTestSq}.  It is important to note that the results of the pressure test will vary with different starting velocities, so we include several representative runs for each case.  As expected, the pressure drops quickly after some amount of time, indicating that these packings are not jammed and that an unjamming motion has been discovered.  Note that more events per sphere are required to observe a pressure ``leak'' when $\phi_c^*-\phi$ is smaller.  In addition, the pressure leak takes slightly longer to show up in the bidisperse case, presumably due to the lessened degree of symmetry in the packing and correspondingly smaller degeneracy of its unjamming motions.

The pop test was also able to demonstrate the existence of ``popping'' motions for all of the aforementioned square lattices.  We also note that the pop test did not perform differently for different jamming gaps, suggesting that it is robust against differences in interparticle distance.

\section{Considerations Regarding Ensemble-Averaging Spectral Density Data}
\label{apx:Fitting}

Great care must be taken when numerically extrapolating $\tilde \chi (k)$ values to $\tilde \chi (0)$.  Additionally, claims about hyperuniformity for small sets of packings with extrapolated $\tilde \chi (0)$ values near zero must be accompanied by reasonable estimates of confidence intervals in order to be valid.  In this Appendix, we show one way in which this might be accomplished.

We mentioned above that the effect of binning data to obtain an ensemble average can give misleading results about the perceived value of $\tilde \chi (k)$, particularly when $\tilde \chi$ is small.  This can be the case when: (i) binning strategies are used to average data from a group of packings produced using the same protocol (but different initial conditions), then (ii) a curve is fit to the binned $\tilde \chi (k)$ values, and finally (iii) the curve is extrapolated to k = 0 to derive $\tilde \chi (0)$. The choice of bin width has a significant effect on the appearance of the ensemble averaged-curve.

For example, considering a set of 100 packings of $N = 2000$ binary disks produced using the LS protocol, binning the $\tilde \chi (k)$ values with bin width equal to $\Delta k=0.01$ and then fitting a 3rd order polynomial for $0 \le k\langle D \rangle / 2\pi \le 0.4$ yields an extrapolated $\tilde \chi (0)$ value of $-4.6 \times 10^{-5}$. However, fitting individual 3rd order polynomials to each packing's $\tilde \chi (k)$ and then averaging the 100 values of $a_0$ yields $\overline{a_0} = 4.6 \times 10^{-5}$.

\begin{figure}[hbt]
  \begin{centering}
    \includegraphics[width=0.4\textwidth,clip]{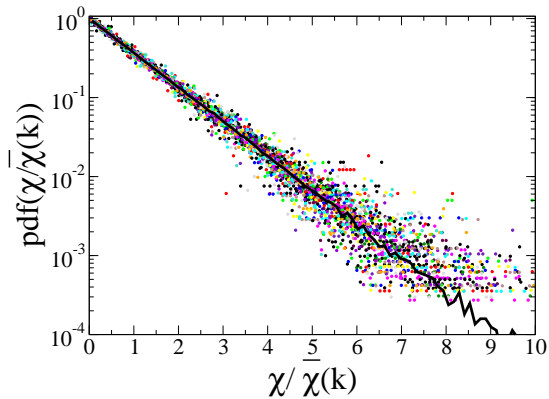}
    \caption{(Color online.) Probability density functions of spectral density values observed in nearly-jammed packings of $N=2000$ bidisperse disks.  One dataset is shown (as a series of filled circles) for each wavenumber, and the thick black line shows the average over all wavenumbers.   Normalizing the distributions with respect to their means makes them collapse to a single master curve, showing that $\tilde \chi(k)$ across an ensemble of packings is approximately exponentially-distributed at any given wavenumber.}
    \label{fig:ChiExpDist}
  \end{centering}
\end{figure}

This difference between a single extrapolated $\tilde \chi (0)$ value and the average $\overline{a_0}$ of individual fits is small but significant. It begs the question, how close to zero must $a_0$ or $\overline{a_0}$ be to be ``effectively hyperuniform''?  To this end, we begin by considering the probability distribution associated with $\tilde \chi(k)$ as a function of the wavenumber $k$.  We take an ensemble of 100 packings of $N=2000$ bidisperse disks and consider the set of spectral densities observed at each wavenumber within the range $0 \le k / 2 \pi \le 0.5$.  From the data at these wavenumbers (normalized by their respective means), probability density functions are obtained; these are shown in Fig.\ \ref{fig:ChiExpDist} as sets of filled circles.  The solid black line is found by aggregating the data together to obtain an average over all wavenumbers.  The data are well-fitted by an exponential distribution, indicating that for wavenumbers $k$ that are near to one another \cite{foot:NotExactlyTheSame} for all of the packings studied, the expected value of any single $\tilde \chi (k)$ is equal to the standard deviation of the distribution.

To determine an estimate for the standard deviation of any extrapolated value $a_0 = \tilde \chi (0)$ for a single packing, we considered the following toy problem: suppose that for a hypothetical $\tilde \chi (k)$ curve, the spectral density corresponding to each wavenumber $0 \le k/ 2 \pi  \le 0.4$ for a hypothetical packing of side length $45 \langle D \rangle$ is chosen from an exponential distribution with a mean given by $\overline{\tilde \chi}(k) = \beta k$, where $\beta$ is chosen empirically to reflect the data from MRJ bidisperse packings. The standard deviation of the constant term in a linear fit to this hypothetical $\tilde \chi (k)$ curve over the range $0 \le k/ 2 \pi \le 0.4$ is $\sigma(a_0) = 4.2 \times 10^{-4}$.  This provides an estimate for the standard deviation of an $a_0$ value extrapolated from the spectral density of any single packing in the aforementioned set of $100$, assuming that the packings in the set exhibit roughly linear behavior in $\tilde{\chi}(k)$ for small $k$ with a near-zero extrapolated $\tilde{\chi}(0)$.

However, the $\overline{a_0}$ value of an ensemble should still converge toward zero as the population of the ensemble grows towards infinity, provided that all of the packings were in fact effectively hyperuniform. To establish to what extent this convergence would occur, we consider a second toy problem in which we generate $20$ sets of $100$ such hypothetical $\tilde \chi (k)$ curves with each $\tilde \chi (k)$ for each curve derived from an exponential distribution as just described. We calculate linear fits and determine $a_0$ for each curve.  We average these values to obtain $\overline{a_0}$ for each set, then compute the mean and standard deviation of these twenty measurements.  We find values of $1 \times 10^{-6}$ and $5 \times 10^{-5}$, respectively.  This suggests that a good estimate for a $\overline{a_0}$ of a set of $100$ packings exhibiting linear and hyperuniform spectral density would be within $\pm 5 \times 10^{-5}$ of zero about 68\% of the time, within $\pm 1.0 \times 10^{-4}$ about 95\% of the time, and within $\pm 1.5 \times 10^{-4}$ about 99\% of the time.

For our $100$ packings with $N=2000$, $\overline{a_0} = 4.6 \times 10^{-5} \pm 6.1 \times 10^{-4}$, falling at about the 64th percentile of the distribution we obtained in the above toy problem (in which the ensemble was exactly hyperuniform by construction), indicating that with about 64\% probability, not all the packings in the set are hyperuniform.  This conclusion suggests that the numerical precision to which we can determine $\tilde{\chi}(0)$ for this set of $100$ packings is not sufficient to rule out effective hyperuniformity with reasonable certainty. A larger sample set might yield more certainty. However, this conclusion supports the notion that jamming is associated with hyperuniformity: only one of the packings in the set passed the pop test and is therefore jammed.

In summary, using the distribution of $\tilde \chi (k)$ values near a given wavenumber $k$, we are able to provide an estimate, for a system of $N=2000$ binary disks produced as described, of a range of extrapolated $\tilde{\chi}(0) = a_0$ values that might be considered hyperuniform. That range is $0.0 \pm 4.2 \times 10^{-4}$, within one standard deviation. Using this previously described method, estimates for confidence intervals over which a set of such packings might be considered effectively hyperuniform can be derived. For $100$ packings of $N=2000$ such disks, an $\overline{a_0}$ value of $0.0 \pm 1.5 \times 10^{-4}$ could be considered hyperuniform. Estimates for larger and smaller sets could be performed as well.

These findings, based on the binning study and the study of the distribution of $\tilde \chi (k)$ values, lead us to suggest that great care must be taken when numerically extrapolating $\tilde \chi (k)$ values to $\tilde \chi (0)$. Additionally, claims about hyperuniformity for small sets of packings with extrapolated $\tilde \chi (0)$ values near zero must be accompanied by reasonable estimates of confidence intervals, perhaps derived from the method described in this Appendix, in order to be valid.

%
%

\ifusebibtex

\bibliography{paper}

\begin{thebibliography}{64}
\bibitem [{foo({\natexlab{b}})}]{foot:spectral}%
  \BibitemOpen
  \bibinfo {note} {The spectral density
  ${\tilde \chi}({\vc k})$ of a two-phase medium is the Fourier transform of
  the autocovariance function, which is trivially related to the probability of
  finding the end points of a vector $\vc r$ inside one of the
  phases.}\BibitemShut {Stop}%
\bibitem [{foo({\natexlab{d}})}]{foot:effective}%
  \BibitemOpen
  \bibinfo {note} {While what one considers
  to be ``effective'' or ``near'' hyperuniformity is ultimately subjective, an
  empirical operational definition for such behavior is that the first peak
  value of the structure factor $S(k)$ [or spectral density ${\tilde \chi}(k)$]
  relative to its value at the origin is roughly $10^{4}$ or larger. In the case of a
  perfect periodic crystal, this ratio diverges to
  infinity.}\BibitemShut {Stop}%
\bibitem [{foo({\natexlab{g}})}]{foot:PopStrict}%
  \BibitemOpen
  \bibinfo {note} {If the simulation box is
  allowed to deform so as to relieve shear stresses, then this test may be used
  to test for strict jamming.}\BibitemShut {Stop}%
\bibitem [{foo({\natexlab{h}})}]{foot:SSDiamReduce}%
  \BibitemOpen
  \bibinfo {note} {The diameters of the
  disks in the soft-sphere packings are all reduced by a constant factor 
  such that no pairs of disks are overlapping and that the closest
  pair of disks are in contact.}\BibitemShut {Stop}%
\bibitem [{foo({\natexlab{i}})}]{foot:DensestRange}%
  \BibitemOpen
  \bibinfo {note} {This is valid for
  $\alpha \in [ \sqrt{2}-1,\sqrt{2}+1 ]$.}\BibitemShut {Stop}%
\bibitem [{foo({\natexlab{j}})}]{foot:NotExactlyTheSame}%
  \BibitemOpen
  \bibinfo {note} {Generally speaking, no
  two packings within an ensemble of MRJ-like packings will have the same
  fundamental cell. Therefore, the wavenumbers at which the spectral density
  can be measured will vary accordingly by a small amount.}\BibitemShut {Stop}%
\end{thebibliography}

\else


%

\fi

\end{document}